# Rough diamond anvils: Steady microstructure, yield surface, and transformation kinetics in Zr

Feng Lin[1]*, Valery I. Levitas[1, 2, 3]*, K. K. Pandey[4], Sorb Yesudhas[1], and Changyong Park[5]

[1]Department of Aerospace Engineering, Iowa State University, Ames, Iowa 50011, USA

[2]Department of Mechanical Engineering, Iowa State University, Ames, Iowa 50011, USA

[3]Ames Laboratory, U.S. Department of Energy, Iowa State University, Ames, Iowa 50011, USA

[4] High Pressure & Synchrotron Radiation Physics Division, Bhabha Atomic Research Centre, Mumbai 400085, India

[5] HPCAT, X-ray Science Division, Argonne National Laboratory, Argonne, Illinois 60439, USA

*Corresponding authors. Email: flin1@iastate.edu and vlevitas@iastate.edu

**Abstract:** Study of the plastic flow and strain-induced phase transformations (PTs) under high pressure with diamond anvils is important for material and geophysics. We introduce rough diamond anvils and apply them to Zr, which drastically change the plastic flow, microstructure, and PTs. Multiple steady microstructures independent of pressure, plastic strain, and strain path are reached. Maximum friction equal to the yield strength in shear is achieved. This allows determination of the pressure-dependence of the yield strength and proves that ω-Zr behaves like perfectly plastic, isotropic, and strain path-independent immediately after PT. Record minimum pressure for α-ω PT was identified. Kinetics of strain-induced PT depends on plastic strain and time. Crystallite size and dislocation density in ω-Zr during PT depend solely on the volume fraction of ω-Zr.

**One-Sentence Summary:** Cell with rough diamond anvils allowed us to determine laws of plastic flow, microstructure, and phase transformations under pressure.

Processes that involve large plastic deformation and PTs under high pressure are common in various manufacturing applications, materials synthesis technologies, and geophysics-related problems. Plastic strain may drastically reduce the PT pressure by one (*1-3*) and even two orders of magnitude (*4*), lead to new phases, and substitute time-controlled kinetics with fast plastic strain-controlled kinetics (*5-8*). Four-scale theory and simulations (*5, 6*) are developed to explain these strain-induced PTs (which are completely different from the traditional pressure or stress-induced PTs). However, it is still in its infancy, and new experimental and theoretical approaches and breakthrough results are very important. The main problem in studying plasticity, plastic strain-induced PTs, and structural changes is that they depend on five components of the plastic strain tensor $\boldsymbol{\varepsilon}_p$ and its entire path $\boldsymbol{\varepsilon}_p^{path}$, making an unspecifiable number of combinations of independent parameters. In particular, the yield surface in the 5D deviatoric stress $\boldsymbol{s}$ space $f(\boldsymbol{s}, \boldsymbol{\varepsilon}_p, \boldsymbol{\varepsilon}_p^{path}) = \sigma_y(p)$ depends on the pressure $p$, $\boldsymbol{\varepsilon}_p$, and $\boldsymbol{\varepsilon}_p^{path}$, demonstrating strain hardening/softening and strain-induced anisotropy (Fig. 1A); here $\sigma_y$ is the yield strength in compression. This complexity makes it practically impossible to determine the complete evolution of the yield surface, even at small strains at ambient pressure. At high pressure, all methods (*9-11*) present the yield surface as $f(\boldsymbol{s}) = \sigma_y(p)$, i.e., like for perfectly plastic material (for which the yield surface is independent of $\boldsymbol{\varepsilon}_p$ and $\boldsymbol{\varepsilon}_p^{path}$, i.e., is fixed in the 5D stress space), and dependence on $\boldsymbol{\varepsilon}_p$ and $\boldsymbol{\varepsilon}_p^{path}$ is neglected and merged in pressure, which causes large error in the determination of the yield strength under high pressure. One of the methods to determine the yield strength in shear $\tau_y = \sigma_y/\sqrt{3}$ in DAC is based on the application of the simplified equilibrium equation $\frac{d\bar{P}}{dr} = -\frac{2\tau_f}{h}$, assuming the contact friction stress $\tau_f$ between anvil and sample reaches its maximum value $\tau_y$ (*10-12*). Here, $\bar{P}$ is the pressure averaged over the sample thickness $h$. However, the results are systematically lower than other methods at ambient and high pressure (*7, 9*) due to the low friction coefficient of diamond leading to $\tau_f < \tau_y$. Coupled simulations and experiments show that $\tau_f = \tau_y$ only in a small region even above 100 GPa (*13*). To resolve the above problems, we introduce *rough diamond anvils (rough-DA)*, whose culet is roughly polished to increase friction (Fig. S1). *We demonstrated that $\tau_f = \tau_y$ for rough-DA, which allowed us to robustly determine $\sigma_y(p)$*. The rough-DA allowed us to solve several other basic problems and brought up discoveries described below.

It was hypothesized in (*12*) that, above some level of accumulated plastic strain $q$ in monotonous straining (straining path without sharp changes in directions), the initially isotropic polycrystalline materials deform as perfectly plastic and isotropic with a strain path-independent surface of the perfect plasticity $\varphi(\boldsymbol{s}) = \sigma_y(p)$ (Fig. 1A). This statement means that the effect of $\boldsymbol{\varepsilon}_p$ and $\boldsymbol{\varepsilon}_p^{path}$ is excluded under the above conditions. Some qualitative supportive arguments for the perfect plastic behavior are presented in (*12*), but quantitative experimental proof is lacking for any material. Here, we heavily pre-deformed commercial Zr by multiple rolling until saturation of its hardness. We show that after the α-ω PT, for four different compression stages (i.e., for very different $\boldsymbol{\varepsilon}_p$ and $\boldsymbol{\varepsilon}_p^{path}$), all pressure distributions in the studied range from 2 to 11 GPa are described by single function $\sigma_y = 1.24 + 0.0965p\ (GPa)$. This is possible only if the *material behaves like perfectly plastic, isotropic, and independent of $\boldsymbol{\varepsilon}_p$ and $\boldsymbol{\varepsilon}_p^{path}$*. The perfectly plastic state is related here to reaching a steady microstructure, determined here by in situ synchrotron X-ray diffraction in terms of crystallite (grain) size $d$ and dislocation density $\rho$, which do not change under successive plastic straining. For rough-DA at the beginning of α-ω

PT, $d_\alpha$ is smaller and $\rho_\alpha$ is larger than those from smooth anvils, i.e., *rough-DA produces different, more refined steady microstructure*. This is also confirmed by the fact that the minimum pressure for plastic strain-induced α-ω PT, $p_\varepsilon^d$, was reduced from 1.36 GPa in smooth DAC to 0.67 GPa with rough DAC, which is the *record low PT pressure* for Zr. For both smooth and rough anvils, the $p_\varepsilon^d$, $d_\omega$, and $\rho_\omega$ in ω-Zr are shown to be independent of $\varepsilon_p$ and $\varepsilon_p^{path}$. Surprisingly, $d_\omega$ and $\rho_\omega$ evolution in ω-Zr during α-ω PT depends solely on the volume fraction $c$ of ω-Zr and is independent of $\varepsilon_p$, $\varepsilon_p^{path}$, $p$, initial $d_\alpha$, and anvil asperities. Similarly, there are unique functions $d_\alpha(c)$ and $\rho_\alpha(c)$ for rough-DA (with some scatter for $0.38<c<0.52$, which is discussed in supplementary materials), independent of $\varepsilon_p$, $\varepsilon_p^{path}$, and $p$. Thus, for strongly pre-deformed material, $\varepsilon_p$ and $\varepsilon_p^{path}$ *are excluded from the governing parameters; this is the main completely unexpected rule for plastic flow, microstructure evolution, and PT under pressure.* The *rough-DA also qualitatively changes the PT kinetics* for $c$: (a) $dc/dq \sim (1-c)$ (first-order reaction) with smooth anvils, while $dc/dq$ is independent of $c$ (zero-order reaction) with rough-DA; (b) In contrast to instantaneous process from conventional view on the strain-induced PTs, $c$ here varies not only with growing plastic strain $q$, but also with time $t$.

*Pressure dependence of the yield strength.* Radial pressure distributions in each phase in five successive compression steps, marked by the peak pressure at the culet center, are shown in Fig. 1B. Corresponding sample thicknesses are collected in Table. S2. Due to the large asperities of the rough-DA, when they penetrate Zr surface, contact sliding occurs in a thin layer of Zr, leading to $\tau_f = \tau_y$. Assuming von Mises yield condition with the yield strength $\sigma_y = \sigma_y^0 + bp$, and taking non-hydrostatic stress and heterogeneity along thickness into consideration, the equilibrium equation averaged over thickness is advanced to (see supplementary materials):

$$\frac{d\bar{P}}{dr} = -A\frac{\sigma_y^0 + b\bar{P}}{h} \to \bar{P} = \left(P_0 + \frac{\sigma_y^0}{b}\right) exp\left(-A\, b\, \frac{r-r_0}{h}\right) - \frac{\sigma_y^0}{b}; \quad A = \frac{2(1+0.524b)}{\sqrt{3}(1-0.262b)}, \qquad (1)$$

where $P_0$ is the pressure at the point $r_0$. Fig. 1C shows that after α-ω PT and for four different compression stages, all pressure distributions overlap and are described by Eq. (1) with single dependence $\sigma_y = 1.24 + 0.0965p\ (GPa)$. Note that $\sigma_y^0 = 1.24\ GPa$ is converted from the hardness of ω-Zr from (*2*), $H$=3.72 GPa, based on the known relationship $\sigma_y^0 = H/3$, proving that $\tau_y$ is reached. Finite element simulations of the processes in DAC (*13, 14*) demonstrate that for different positions and compression stages, $\varepsilon_p$, $\varepsilon_p^{path}$, and material rotations are very different. Consequently, the ability to describe all four curves with single function $\sigma_y(p)$ demonstrates strict proof that for the monotonous loading with rough-DA, ω-Zr deforms as perfectly plastic and isotropic material with $\varepsilon_p$ and $\varepsilon_p^{path}$ independent surface of perfect plasticity. Additional point is that the perfectly plastic state is found almost immediately after completing α-ω PT, i.e., it is inherited from α-Zr. We found that for smooth anvils up to *15 GPa*, the ratio $\tau_f/\tau_y = 0.39\text{-}0.46$ away from the center characterizes underestimate in the $\sigma_y(p)$ in previous works. We connect perfectly plastic behavior with reaching steady microstructure. After completing PT, $d_\omega$, and $\rho_\omega$ for 6, 10, and 14 GPa steps are practically independent of radius (Figs. 2B and 3B). Since $\varepsilon_p$, $\varepsilon_p^{path}$, and $p$ strongly vary with radius and increasing load, this indicates that steady microstructure, which is independent of pressure, $\varepsilon_p$, and $\varepsilon_p^{path}$, is reached.

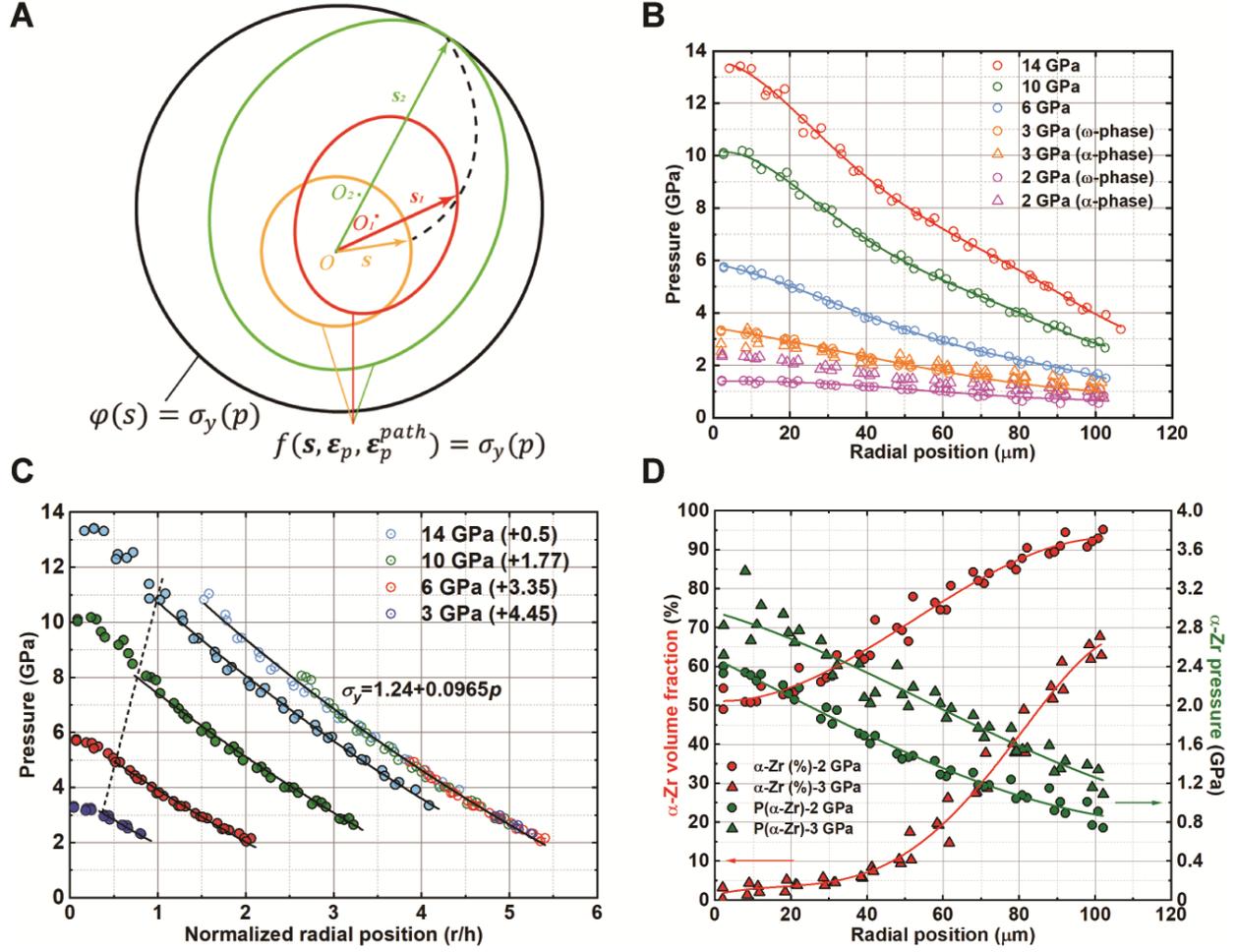

**Fig. 1. Determination of the surface of perfect plasticity for ω-Zr.** (**A**) Schematic of the evolution of the yield surface $f(\mathbf{s}, \boldsymbol{\varepsilon}_p, \boldsymbol{\varepsilon}_p^{path}) = \sigma_y(p)$ until it reaches the fixed surface of perfect plasticity $\varphi(\mathbf{s}) = \sigma_y(p)$ in "5D" space of deviatoric stresses $\mathbf{s}$ at fixed $p$. The initial yield surface and $\varphi(\mathbf{s}) = \sigma_y(p)$ are isotropic (circles). Two other yield surfaces depend on $\boldsymbol{\varepsilon}_p$ and $\boldsymbol{\varepsilon}_p^{path}$, and acquire strain-induced anisotropy, namely shifted centers $O_1$ and $O_2$ (back stress) and ellipsoidal shape due to texture. When the yield surface reaches $\varphi(\mathbf{s}) = \sigma_y(p)$, the material deforms as perfectly plastic, isotropic with the fixed surface of perfect plasticity. (**B**) Pressure distributions for different deformation steps with rough-DA. (**C**) Pressure in single-phase ω-Zr vs. r/h. Solid lines correspond to Eq. (1) for $\sigma_y^0 = 1.24\ GPa$ and b=0.0965. Eq. (1) is not valid around culet center. Dash line shows the position where data is truncated. The unified curve for all loadings (necessary to use data from all four compression stages as a single data set) is obtained by shifting each curve (which is allowed by differential Eq. (1), see supplementary materials) along the horizontal axis by an appropriate distance. Shifts are shown in parenthesis. Since for different points from different curves $\boldsymbol{\varepsilon}_p$, $\boldsymbol{\varepsilon}_p^{path}$, and material rotations are very different, the obtained results prove the perfectly plastic and isotropic material response with $\boldsymbol{\varepsilon}_p$ and $\boldsymbol{\varepsilon}_p^{path}$ independent surface $\varphi(\mathbf{s}) = \sigma_y(p)$. (**D**) Distribution of volume fraction of α-Zr and pressure in α-Zr. Starting from the 6 GPa step, α-Zr is fully transformed to ω-Zr along the radius. Note that errors from the Rietveld refinement of the x-ray patterns for pressure, the volume fraction of phases (as well as dislocation density and crystallite size) are smaller than the symbols in the plots.

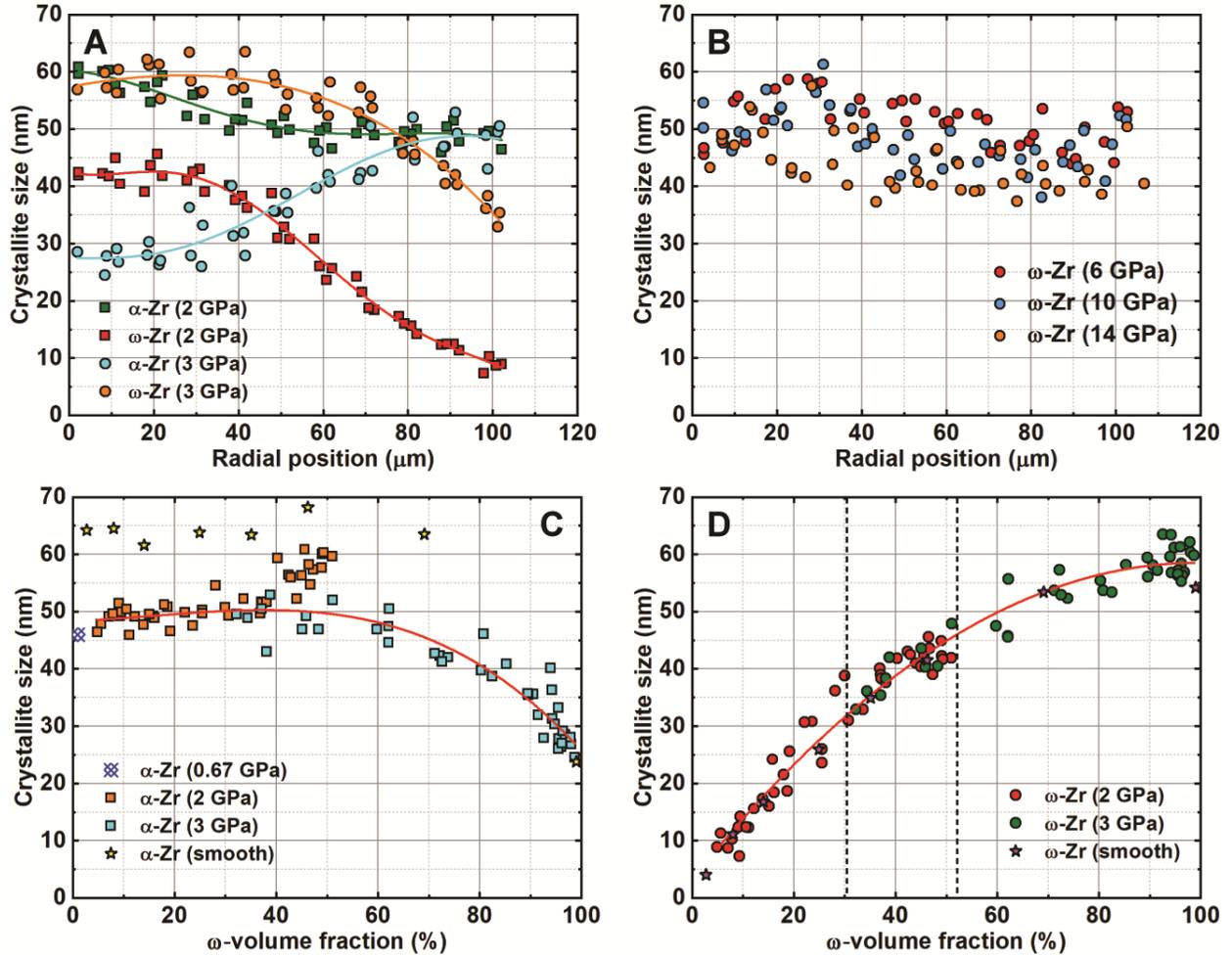

**Fig. 2 Crystallite size distribution and evolution.** Radial distributions of the crystallite size obtained with rough-DA at (**A**) 2 and 3 GPa steps and (**B**) 6, 10, and 14 GPa steps. Since $\varepsilon_p$, $\varepsilon_p^{path}$, and $p$ strongly vary with radius and increasing load, this indicates that steady microstructure in terms of crystallite size, which is independent of pressure, $\varepsilon_p$, and $\varepsilon_p^{path}$, is reached almost immediately after PT. (**C**) Crystallite size of α-Zr versus steady volume fraction of ω-Zr from 2 and 3 GPa steps with rough-DA and with smooth anvils. Blue cross represents the first appearance of ω-Zr with rough-DA at 0.67 GPa. For rough-DA at the beginning of α-ω PT, the crystallite size is smaller than that from smooth anvils, i.e., rough-DA produces different, more refined steady microstructure. With exception of region 0.38<c<0.52, where some scatter is observed, the crystallite size of α-Zr during α-ω PT is the unique function of $c$, almost constant for c<0.6, which is independent of pressure, plastic strain, and strain path. (**D**) Crystallite size of ω-Zr versus steady volume fraction of ω-Zr from 2 and 3 GPa steps with rough-DA, and with smooth anvils. For rough-DA, points from 2 and 3 GPa steps overlap within dash lines. Results in (D) represent surprising rule for ω-Zr for both rough and smooth anvils: existence of the unique curve for the crystallite size solely depending on $c$ for both pressure steps during α-ω PT independent of pressure, $\varepsilon_p$ and $\varepsilon_p^{path}$.

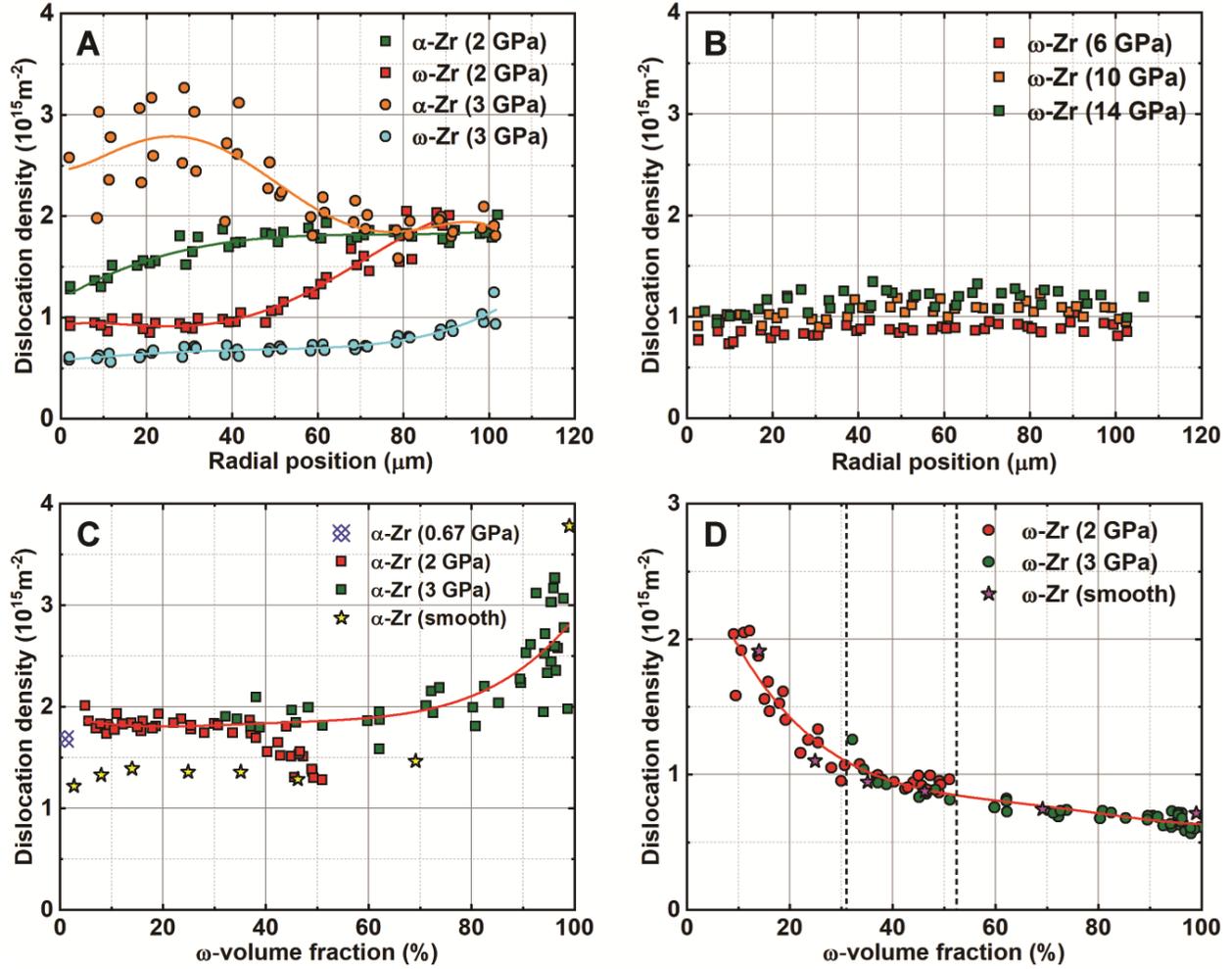

**Fig. 3. Dislocation density distribution.** Radial distribution of dislocation density at (**A**) 2 and 3 GPa steps, and (**B**) 6, 10, and 14 GPa steps. Since $\varepsilon_p$, $\varepsilon_p^{path}$, and $p$ strongly vary with radius and increasing load, this indicates that steady microstructure in terms of dislocation density, which is independent of pressure, $\varepsilon_p$, and $\varepsilon_p^{path}$, is reached almost immediately after PT. (**C**) Dislocation density in α-Zr versus volume fraction of ω-Zr from 2 and 3 GPa steps with rough-DA, and with smooth anvils. Blue cross represents the first appearance of ω-Zr with rough-DA at 0.67 GPa. Since for rough-DA at the beginning of α-ω PT, the dislocation density is larger than from smooth anvils, the rough-DA produces different, more defected steady microstructure. With exception in region 0.38<c<0.52, where some scatter is observed, the dislocation density in α-Zr during α-ω PT is the unique function of c, almost constant for c<0.6, which is independent of pressure, plastic strain, and strain path. (**D**) dislocation density in ω-Zr versus volume fraction of ω-Zr from 2 and 3 GPa steps with rough-DA, and with smooth anvils. For rough-DA, points from 2 and 3 GPa steps overlap within dash lines. Results in (D) represent unexpected law for ω-Zr for both rough and smooth anvils: existence of the unique curve for the dislocation density solely depending on $c$ for both pressure steps during α-ω PT independent of pressure, $\varepsilon_p$ and $\varepsilon_p^{path}$. For small $c$, $\rho_\alpha = \rho_\omega$, indicating that small nuclei directly inherit the dislocation structure from α-Zr during strain-induced PT.

*Minimum pressure for initiation of strain-induced PT $p_\varepsilon^d$.* ω-Zr diffraction peaks started being observed at $p_\varepsilon^d = 0.67$ GPa at the sample center (Figs. 4A and S2). This is a record low pressure for α-ω Zr PT, which is 9.0 times lower than that under hydrostatic loading ($p_h^d = 6.0$ GPa), 5.1 times lower than the phase equilibrium pressure of 3.4 GPa (*15*), and 2 times lower than $p_\varepsilon^d = 1.36$ GPa obtained with smooth anvils. At the culet edge at 2 GPa step, c=0.05 at 0.74 GPa (Fig. 1D), which means $p_\varepsilon^d$ at the edge is practically identical to that at the center. This indicates that for strongly pre-deformed α-Zr, $p_\varepsilon^d$ is independent of $\varepsilon_p$, $\varepsilon_p^{path}$ and pressure-strain path since they are very different at center and edge. The same is true for smooth diamonds (Fig. S3), for which, due to higher PT pressure, we have more points with $p = p_\varepsilon^d$.

At the initiation of PT with rough-DA, $d_\alpha \approx 46$ nm (Fig. 3C) and $\rho_\alpha = 1.68 \times 10^{15}/m^2$ (Fig. 4C), while with smooth anvils, $d_\alpha \approx 66$ nm (Fig. 3C) and $\rho_\alpha = 1.22 \times 10^{15}/m^2$ (Fig. 4C), both independent of radii and, consequently, $\boldsymbol{\varepsilon}_p$, $\boldsymbol{\varepsilon}_p^{path}$ and pressure-strain path. Thus, smooth and rough anvils produce different steady microstructures in α-Zr, which results in different $p_\varepsilon^d$. For steady microstructure of pre-deformed α-Zr, d≈74 nm and $\rho = 9.94 \times 10^{14}/m^2$ at ambient condition, which is one more steady microstructure. Since for annealed α-Zr with micron grains, $p_\varepsilon^d = 2.3\ GPa$ (7), a general trend is that $p_\varepsilon^d$ reduces with reduction in $d_\alpha$ (opposite to the initial theoretical prediction in (5)) and increase in $\rho_\alpha$.

*PT and microstructure evolution kinetics.* Distributions of pressure in phases and volume fraction $c$ of ω-Zr at 2 and 3 GPa steps are presented in Fig. 1D. Strain-induced PT kinetic equation derived based on nanoscale mechanisms (5) with neglected reverse PT is:

$$\frac{dc}{dq} = k \frac{B(1-c)^a}{B(1-c)+c} \left(\frac{p_\alpha(q) - p_\varepsilon^d}{p_h^d - p_\varepsilon^d}\right) \quad \text{for} \quad p_\alpha > p_\varepsilon^d. \tag{2}$$

Here $p_\alpha(q)$ is the pressure in α-Zr - $q$ loading path; $B = \left(\frac{\sigma_y^\omega}{\sigma_y^\alpha}\right)^l$; $k$ and $l$ are material parameters. For smooth anvils, $a=1$, $k=11.65$, and $B=1.35$ (Fig. S4). Although plastic strain tensor at arbitrary $r$ is unknown in experiments, material near the symmetry axis undergoes uniaxial compression and $q = ln(h_0/h)$. Through numerical integration of Eq. (2), $c$ can be expressed as a function of $I = \int_{q_0}^{q} (p_\alpha(q) - p_\varepsilon^d)\ dq$, where $q_0$ is the accumulated plastic strain at $p_\varepsilon^d$. In addition to steady-state data (after long relaxation time), data instantly after compression and transient data between instant and steady states are shown in Fig. 4B. Such an unexpected time dependence of PT kinetics confronts the conventional view that strain-induced PTs do not occur without plastic strain increment, time is not an essential parameter, and plastic strain serves as time-like parameter (like in Eq. (2)) (5, 6, 8). Note that since the thickness of the sample does not change between instant and steady states, creep as a reason for the time dependence of the strain-induced PT is excluded. It appears that rough-DA allows us not only to reveal the time-dependent part of the growth for strain-induced PT, but also to change the plastic strain-dependent part. Surprisingly, $c$-$I$ curve is linear for steady state and instant state before relaxation at 2 GPa step ($I < 0.5$) and after relaxation, with practically the same slope (Fig. 4B). Thus, the rate of PT in Eq. (2) is independent of $c$, which results in $a=l=0$, $B=1$, and

$$\frac{dc}{dq} = k \frac{p_\alpha(q) - p_\varepsilon^d}{p_h^d - p_\varepsilon^d} \tag{3}$$

Value $a=1$ corresponds to multiple nucleation within the parent phase, while $a=0$ is typical for propagation from a limited number of nuclei without their interaction, like for thickening of PT band. Eq. (3) should be used for each fast-loading increment and for steady state, with different $k$. Time-dependent contribution to the kinetics that reproduces Eq. (3) for the instant kinetics at $t = 0$ and steady-state kinetics for $t = \infty$ and describes transient data at 2 GPa step is:

$$c(t) = c(q)_{t=\infty} + \left(c(q)_{t=0} - c(q)_{t=\infty}\right) \exp\left(-\frac{t}{43.13}\right) \tag{4}$$

with a characteristic time of 43.13 min. Here, $c(q)_{t=0}$ and $c(q)_{t=\infty}$ are the volume fractions after instant compression and in the steady state. The surprising rule is found in Figs. 2D and 3D for ω-Zr: the unique curves $d_\omega(c)$ and $\rho_\omega(c)$ for both pressure steps during α-ω PT independent of pressure, $\varepsilon_p$ and $\varepsilon_p^{path}$; for $d_\omega(c)$, it is also independent of processing with rough and smooth anvils. There is similar, but weaker regularity for $d_\alpha(c)$ and $\rho_\alpha(c)$ (Figs. 2C and 3C): with exception in region $0.38<c<0.52$, where some scatter is observed (see supplementary material), the crystallite size and dislocation density in α-Zr during α-ω PT is the unique function of $c$, almost constant for $c<0.6$, which is independent of pressure, plastic strain, and strain path. For small $c$, $\rho_\alpha =\rho_\omega$, indicating that small nuclei directly inherit the dislocation structure from α-Zr during strain-induced PT.

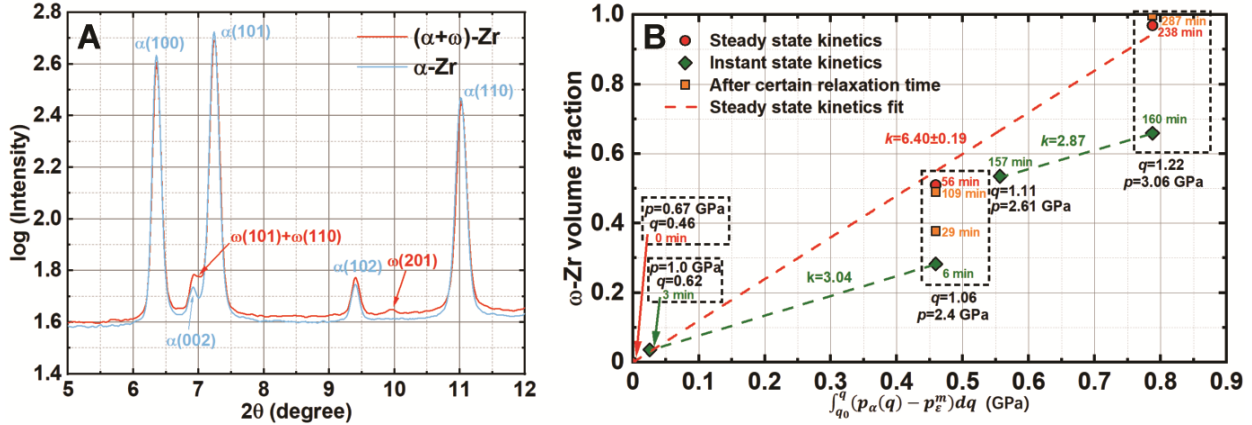

**Fig. 4 First appearance of ω-Zr at the sample center and strain and time-dependent kinetics of α-ω PT in Zr.** (**A**) Pure α-Zr diffraction peaks (blue) at $p=$ 0.49 GPa and appearance of ω-Zr peaks at $p_\varepsilon^d = 0.67$ GPa (red). (**B**) The volume fraction of ω-Zr vs. plastic strain and time. Green diamonds represent diffraction data after instant compression; red circles designate results after reaching steady state; orange squares show intermediate data vs. relaxation time; time labels show time from the beginning of the first measurement; $q$-values are shown near symbols. Revealed linear strain-dependent kinetics and time dependence of the kinetics of strain-induced PT were not observed in the literature.

The current results not only present the main and very nontrivial rules of plastic yielding, strain-induced PT, and microstructure evolution during and after PT under high pressure but also open new windows for utilizing rough-DA and finding similar laws for multiple material systems in a broad pressure range. In particular, one can determine the pressure dependence of the yield strength for important multiphase material systems (e.g., mantle rocks and composites). Discovered time-dependence of the kinetics of strain-induced PTs opens unexplored field of the simultaneous strain- and stress-induced PTs under pressure. By optimizing anvil asperity, desirable plastic flow, minimum grain size, and minimum PT pressure can be reached. Also, instead of severe plastic straining at high pressure, e.g., by high-pressure torsion, one can reach one of the steady microstructures by severe straining at normal pressure (e.g., by rolling, ball milling, or equal channel extrusion) and then produce PT and reach steady microstructure with smaller grain size at relatively small plastic strain and low pressure by compression or high-pressure torsion. Holding at a constant load to utilize the time-dependent PT component may also be useful. For small volume fraction of ω-Zr, crystallite size is much smaller, and dislocation density is larger than for the steady state. This gives an idea of designing α-ω Zr composites with increased strength due to strong ultrafine-grained ω-Zr and sufficient plasticity due to α-Zr. During intense loading, an increase in volume fraction of ω-Zr leads to energy absorption and an increase in strength. All these may result in the economic plastic strain-induced synthesis of nanostructured high-strength high-pressure phases at low pressures. In addition, rough-DA eliminates the problem of describing contact friction required for modeling deformational and PT processes in DAC (*13, 14, 16*). For traditional high-pressure torsion with ceramic/metallic anvils, friction reaches the maximum possible level due to large asperities. Utilizing rough-DA in rotational DAC (*1, 4, 6*) will allow in situ studies of high-pressure torsion. Also, to increase the maximum possible pressure in DAC, toroidal grooves are used, which increase friction (*17*). This can be done with rough-DA more uniformly throughout the culet and with smaller stress concentrators. Note that the above plethora of results was obtained in a single experiment, thus transforming the main challenge—strongly heterogeneous fields—into a great opportunity.

**Acknowledgments:** The authors thank Drs. Alexander Zhilyaev and María Teresa Pérez-Prado for providing the same type of Zr sample as they used in (*2*). The authors also thank Dr.

Reinhard Boehler for preparing the surface of rough-DA. The help of Prof. Ashraf Bastawros and Dr. Bishoy Dawood in measuring the asperity profile of the culets is greatly appreciated.

**Funding:** Support from NSF (CMMI-1943710 and DMR-1904830), and Iowa State University (Vance Coffman Faculty Chair Professorship) is greatly appreciated. This work is performed at HPCAT (Sector 16), Advanced Photon Source (APS), Argonne National lab. HPCAT operations are supported by DOE-NNSA's Office of Experimental Science. The Advanced Photon Source is a U.S. Department of Energy (DOE) Office of Science User Facility operated for the DOE Office of Science by Argonne National Laboratory under Contract No. DE-AC02-06CH11357.

**Author contributions:** FL and KKP performed experiments, collected and postprocessed data. VIL conceived the study, supervised the project, and developed theoretical models. SY performed experiments. CP assisted with experiments. FL and VIL prepared initial manuscript. All authors contributed to discussions of the data and to the writing of the manuscript.

## Competing interests

The authors declare no competing interests.

## Data availability

The data that support the findings of this study are available from the corresponding authors upon request.

# Supplementary Materials for

## Rough diamond anvils: Steady microstructure, yield surface, and transformation kinetics in Zr

Feng Lin[1]*, Valery I. Levitas[1,2,3]*, K. K. Pandey[4], Sorb Yesudhas[1], and Changyong Park[5]

[1]Department of Aerospace Engineering, Iowa State University, Ames, Iowa 50011, USA

[2]Department of Mechanical Engineering, Iowa State University, Ames, Iowa 50011, USA

[3]Ames Laboratory, U.S. Department of Energy, Iowa State University, Ames, Iowa 50011, USA

[4] High Pressure & Synchrotron Radiation Physics Division, Bhabha Atomic Research Centre, Mumbai 400085, India

[5] HPCAT, X-ray Science Division, Argonne National Laboratory, Argonne, Illinois 60439, USA

*Corresponding authors. Email: flin1@iastate.edu and vlevitas@iastate.edu

**This PDF file includes:**

Materials and Methods
Figs. S1 to S6
Tables S1 to S2
Supplementary Text
References

**Materials and Methods**

Starting materials and experiment details

The material studied here is the same as was used in (*18*), purchased from Haines and Maassen (Bonn, Germany), which is commercially pure (99.8%) alpha Zr (Fe: 330 ppm; Mn: 27 ppm; Hf: 452 ppm; S: <550 ppm; Nd: <500 ppm). The sample slab of the initial thickness of 5.25 mm was cold-rolled down to thin foil to obtain a plastically pre-deformed sample with saturated hardness. 3 mm disk was punched out from thin foil for unconstrained non-hydrostatic compression experiments in DAC. For hydrostatic compression experiments, specks of size 20 µm were chipped off from the pre-deformed sample. The hydrostatic high-pressure x-ray diffraction measurements were performed to constrain the 3$^{rd}$ order Birch-Murnaghan equation of state (Table S1) and pressure $p_h^d$, which was found to be 6.0 GPa. All the pressures in unconstrained non-hydrostatic experiments are determined using measured lattice parameters and cell volume with the same equation of state. For hydrostatic experiments, Zr specks of size 20 µm were loaded in the sample chamber along with silicone oil and copper chips as pressure transmitting medium and pressure marker, respectively. The sample chamber was prepared by drilling a hole of 250 µm diameter in pre-indented stainless-steel gaskets indented from the initial thickness of 250 µm to 50 µm. Hydrostatic high-pressure experiments were carried out in a small pressure step of 0.2 GPa up to a maximum pressure of 16 GPa. For the non-hydrostatic experiment with smooth diamond anvils, a pre-deformed Zr sample disk (3 mm diameter, initial thickness 165 µm) was gradually compressed to ~15 GPa at the culet center without any constraining gasket using a custom-designed loading system. For the nonhydrostatic experiments with rough diamond anvil (rough-DA) (Fig. S1), a pre-deformed Zr disk sample (3 mm diameter, initial thickness 163 µm) was compressed gradually up to ~14 GPa at the culet center with a gas-membrane system.

All the in-situ axial XRD experiments were performed at 16-BM-D beamline at HPCAT (Sector 16) at Advanced Photon Source employing focused monochromatic x-rays of wavelength 0.3100 Å and size of 6*µm* x 5*µm* (full width at half maximum (FWHM)) and recorded with Perkin Elmer detector. For the smooth anvil experiment, the sample was scanned along one culet diameter (500 µm) in 10 µm step size at each load. For rough-DA experiment, the sample was scanned along two perpendicular culet diameters (230 µm) in 10 µm step size. The sample thickness was measured through x-ray intensity absorption using the linear attenuation equation with density corrected to the corresponding pressure, similar to (*7*). For the rough-DA experiment, the thickness was measured for six steps shown in Table S2: when ω-Zr emerged (0.67 GPa at the center) and when the pressure at the center reached ~2, 3, 6, 10, and 14 GPa. The diffraction images were first converted to unrolled patterns using FIT2D software (*19*) (Fig. S2) and then analyzed through Rietveld refinement using MAUD software (*20*) to obtain the lattice parameters, volume fractions of ω-Zr, microstrains, and crystallite sizes.

Dislocation density estimation

The crystallite sizes and microstrains extracted from the refinement using MAUD were used to estimate the dislocation density as well, which helps in situ tracking the microstructure change during deformation. Dislocation density can be expressed as (*21*):

$$\rho = \sqrt{\rho_c \rho_{ms}} \ . \qquad (S1)$$

Where $\rho_c$ and $\rho_{ms}$ are the contribution to overall dislocation density from crystallite size and microstrain, respectively. Contribution from crystallite size is:

$$\rho_c = \frac{3}{d^2}. \tag{S2}$$

Where $d$ is the crystallite size. Contribution from the microstrain is determined by the equation:

$$\rho_{ms} = k\varepsilon^2/b^2. \tag{S3}$$

Where $\varepsilon$ is the microstrain; $b$ is the magnitude of the Burgers vector; $k = 6\pi A(\frac{E}{G \ln(r/r_0)})$ is a material constant; $E$ and $G$ are Young's modulus and shear modulus, respectively; $A$ is a constant that lies between 2 and $\pi/2$ based on the distribution of strain; $r$ is the radius of crystallite with dislocation; $r_0$ is a chosen integration limit for dislocation core. In this study, $A = \pi/2$ as the gaussian distribution of strain. Moduli $E$, $G$ and their pressure dependence for α and ω-Zr are taken from (22) and (23), respectively. A reasonable value of $\ln(r/r_0)$ being 4 is used (21). α-Zr has a dominant prismatic slip system of $\{1\bar{1}00\}\langle11\bar{2}0\rangle$ (e.g., 24-27). As for ω-Zr, (28) suggests a prismatic $\{11\bar{2}0\}\langle1\bar{1}00\rangle$ and basal $\{0001)\}\langle1\bar{1}00\rangle$ dominant slip system based on plasticity modeling. Since crystal lattice gets compressed under pressure, the length of the Burger vector is calculated using pressure-dependent lattice constants. It is worthy to note that when estimating dislocation density using the Williamson-Smallman method (21), we only consider one dominant dislocation slip system. However, to accommodate arbitrary imposing plastic strain on polycrystal, auxiliary slip systems are usually needed. With changing orientation of grains during deformation, the Schmid factor of slip systems changes, and thus slip system activities, which is the percentage of plastic strain accommodated by certain slip systems, will be different. This will induce uncertainty in dislocation density estimation.

Evaluation of the yield strength under high pressure

Pressure dependence of the yield strength is of great interest to many disciplines for various reasons. It determines: (a) strength of structural elements working under extreme loads, in particular, different high-pressure apparatuses, including DAC, rotational DAC, and apparatuses with metallic or ceramic dies for the high-pressure torsion; (b) maximum pressure that can be achieved in materials compressed in DAC (see Eq. (1)); (c) material flow in different technologies, like high-pressure material synthesis, extrusion, forging, cutting, polishing, and ball milling; (d) maximum possible friction in heavily loaded contacts, and related wear; (e) the level of shear (deviatoric) stresses that can be applied to materials. The shear stresses drastically affect the phase transformations, chemical reactions, and other structural changes (1-8); (f) plastic flow and geodynamic processes in Earth and other planets, including earthquakes.

There are two approaches to estimate yield strength under pressure in a DAC-like device, which exploit x-ray diffraction in either radial or axial diffraction geometry. With radial diffraction geometry, the yield strength in compression can be estimated from the lattice strains (distortion of crystal lattice planes) measured by synchrotron x-ray diffraction. Since the compression direction is perpendicular to the x-ray beam, lattice strains are detectable because axial compression symmetry and diffraction symmetry do not coincide. With this method, all the components of the elastic strain tensor in single crystals comprising polycrystalline sample can be determined. Combined with high-pressure single crystal elastic constants, lattice strains can be used to estimate the yield strength with proper mechanical assumptions (29). Despite obtaining a large amount of experimental information and broad usage, this method suffers from several disadvantages:

(a) All measurements are averaged over the diameter of the sample, and the radial gradient of strain and stress fields is unavoidable due to contact friction. The macroscopic stress state also includes shear stresses, which are not included in the treatment. To reduce the effect of friction, a relatively small ratio of the sample diameter to thickness $d/h$ needs to be used, which also limits the axial displacement and applied plastic strain.

(b) When estimating yield strength from the lattice strains, different chosen mechanical assumptions to determine effective elastic properties of the polycrystalline aggregate (Reuss, Voigt, Hill, self-consistent, etc.) leads to different results.

(c) For multiphase materials, lattice strains give an estimation of stress in a single phase only. The mixture theory for the yield strength of multiphase material is not well developed, especially for large difference in the yield strength of phases (*30, 31*).

(d) Yield strength depends on the pressure, plastic strain, and grain size that evolve during deformation. By presenting the yield strength versus pressure, all these effects are prescribed to the pressure only, which introduces large errors.

With axial diffraction geometry, yield strength is estimated using radial pressure gradient and sample thickness based on the simplified mechanical equilibrium equation in radial direction $r$ (*10-12*), combined with the assumption that the friction stress reaches the yield strength in shear $\tau_y$:

$$\frac{d\bar{p}}{dr} = -\frac{2\tau_y(p)}{h}, \tag{S4}$$

where $\bar{p}$ is the pressure, averaged over the sample thickness. Previously, the pressure was measured at the surface using the ruby fluorescence method and thickness was measured on recovered samples after unloading. Currently, pressure $\bar{p}$ can be measured using x-ray diffraction and thickness using x-ray absorption. The advantage of Eq. (S4) is that it does not include constitutive equations and assumptions, making it available for multiphase material. Disadvantages are:

(a) Due to the low friction coefficient of diamond, the friction stress is much lower than the yield strength in shear $\tau_y$. This is the reason why this method significantly underestimates the yield strength.

(b) Stress $\boldsymbol{\sigma}$ and strain $\boldsymbol{\varepsilon}_p$ tensor fields are strongly heterogeneous along the radius, and material undergoes very different plastic straining path $\boldsymbol{\varepsilon}_p^{path}$ at different positions. Since the yield strength depends on pressure, $\boldsymbol{\varepsilon}_p$, and $\boldsymbol{\varepsilon}_p^{path}$, but is presented as a function of a pressure only, this also introduces large errors.

(c) Eq. (S4) neglects heterogeneity along the thickness and difference between pressure and normal stresses.

We eliminate all the above drawbacks and advance mechanical equilibrium Eq. (S4) to the form of Eq. (1) from the main text, which considers the heterogeneity of all stresses across the sample thickness, in the following part.

**Supplementary Text**
**Derivation of the advanced averaged equilibrium equation**

*Problem formulation.* For compression of a sample in the DAC, $\sigma_{33}$, $\sigma_{11}$, and $\sigma_{22}$ are the normal stress components along the load (vertical), radial, and azimuthal directions, respectively; $\tau_{31}$ is the shear stress; $\sigma_y$ and $\tau_y$ are the yield strength in compression and shear respectively. Compressive stresses are negative. Pressure is defined as:

$$p = -(\sigma_{11} + \sigma_{22} + \sigma_{33})/3. \tag{S5}$$

All stresses and pressure are functions of $r$ and $2z/h$ in a cylinder coordinate system with the origin at the center of the sample cylinder, where $h$ is the sample thickness; in particular, $p(0)$ corresponds to the symmetry plane $z = 0$ and $p(1)$ corresponds to the contact surface $2z/h = 1$. Pressure (or any stress), averaged over the sample thickness, is defined as:

$$\bar{p} = \frac{1}{h}\int_0^h p \, dz. \tag{S6}$$

The contact friction stress $\tau_f$ is defined by the simplified mechanical equilibrium equation

$$\frac{d\bar{\sigma}_{11}}{dr} = -\frac{2\tau_f(p(1))}{h}. \tag{S7}$$

The pressure-dependent yield strength in compression $\sigma_y$ and shear $\tau_y = \sigma_y/\sqrt{3}$ (based on the von Mises equivalent stress) are:

$$\sigma_y = \sigma_y^0 + bp; \qquad \tau_y = \sigma_y/\sqrt{3} = (\sigma_y^0 + bp)/\sqrt{3}. \tag{S8}$$

Note that $\sigma_y$ depends on the local pressure $p$. At the contact surface, symmetry plane, and for averaged over the thickness, we have different pressures and yield strengths:

$$\sigma_y(1) = \sigma_y^0 + bp(1); \quad \sigma_y(0) = \sigma_y^0 + bp(0); \quad \bar{\sigma}_y = \sigma_y^0 + b\bar{p}. \tag{S9}$$
$$\tau_y(1) = \left(\sigma_y^0 + bp(1)\right)/\sqrt{3}; \quad \tau_y(0) = \left(\sigma_y^0 + bp(0)\right)/\sqrt{3}; \quad \bar{\tau}_y = \left(\sigma_y^0 + b\bar{p}\right)/\sqrt{3}.$$

For maximum possible friction provided by the rough-DA we have:

$$\tau_f(p(1)) = \tau_y(1) = \frac{1}{\sqrt{3}}\sigma_y(1) = \frac{1}{\sqrt{3}}\left(\sigma_y^0 + bp(1)\right). \tag{S10}$$

With expression in Eq. (S10), the equilibrium Eq. (S7) specifies as:

$$\frac{d\bar{\sigma}_{11}}{dr} = -\frac{2}{\sqrt{3}}\frac{\sigma_y(1)}{h} = -\frac{2}{\sqrt{3}}\frac{\sigma_y^0 + bp(1)}{h}. \tag{S11}$$

Since we assume that in XRD experiments, the distribution of pressure $\bar{p}(r)$ averaged over the thickness is measured, we need to express $\bar{\sigma}_{11}$ and $p(1)$ in Eq. (S11) in terms of $\bar{p}(r)$. Traditionally, this difference is neglected, i.e., it is assumed $\bar{\sigma}_{11} = p(1) = \bar{p}(r)$, which introduces errors.

*Analytical evaluation of the stress and pressure fields.* We assume that material behaves as perfectly plastic and isotropic macroscopically, with the surface of perfect plasticity $\varphi(\boldsymbol{s}) = \sigma_y(p)$ in the 5D deviatoric stress tensor $\boldsymbol{s}$ space. This surface is independent of the plastic strain tensor $\boldsymbol{\varepsilon}_p$ and its path $\boldsymbol{\varepsilon}_p^{path}$. Such behavior can be achieved after large enough preliminary plastic deformation leading to saturation of hardness (*12*). The pressure-dependent von Mises yield condition (i.e., Drucker-Prager yield condition) is assumed:

$$\varphi(\boldsymbol{s}) = \frac{1}{\sqrt{2}}\sqrt{(\sigma_{11}-\sigma_{22})^2 + (\sigma_{11}-\sigma_{33})^2 + (\sigma_{22}-\sigma_{33})^2 + 6\tau_{13}^2} = \sigma_y(p) = \sqrt{3}\tau_y(p). \tag{S12}$$

Equilibrium equations are:

$$\frac{\partial\sigma_{11}}{\partial r} + \frac{\partial\tau_{13}}{\partial z} + \frac{\sigma_{11}-\sigma_{22}}{r} = 0; \tag{S13}$$

$$\frac{\partial\sigma_{33}}{\partial z} + \frac{\partial\tau_{13}}{\partial r} + \frac{\tau_{13}}{r} = 0. \tag{S14}$$

The following assumptions are made:
(a) It approximately follows from the finite element method simulations and DAC experiments: $\sigma_{11} = \sigma_{22}$. Then plasticity condition Eq. (S12) simplifies to:

$$(\sigma_{11}-\sigma_{33})^2 + 3\tau_{31}^2 = \sigma_y^2(p) = 3\tau_y^2(p). \tag{S15}$$

(b) Stress $\sigma_{33}$ is independent of $z$. However, it does not mean that:

$$\frac{\partial\tau_{13}}{\partial r} + \frac{\tau_{13}}{r} = 0 \quad \rightarrow \quad \tau_{13} = \tau_0(z)\frac{r_0}{r}. \tag{S16}$$

because at the contact surface, $\tau_0(z)$ may equal to constant $\sigma_y$ for all $r$, for material with pressure-independent yield strength. $\sigma_{33}$, that is independent of $z$ means two other terms in Eq. (S14) make small contributions to $\sigma_{33}$.

For plane strain, when the term $\frac{\tau_{13}}{r}$ in Eq. (S14) is absent, a slightly modified Prandtl's solution for the maximum possible contact friction (*32*) for stresses that satisfy equilibrium equations and plasticity conditions are:

$$\frac{\sigma_{33(r)}}{\tau_y} = \frac{\sigma_{33}(0)}{\tau_y} + \frac{2r}{h};  \tag{S17}$$

$$\frac{\tau_{13}}{\tau_y} = \frac{2z}{h};  \tag{S18}$$

$$\frac{\sigma_{11}}{\tau_y} = \frac{\sigma_{33}(0)}{\tau_y} + \frac{2r}{h} + \sqrt{3}\sqrt{1-\left(\frac{2z}{h}\right)^2} = \frac{\sigma_{33(r)}}{\tau_y} + \sqrt{3}\sqrt{1-\left(\frac{2z}{h}\right)^2};  \tag{S19}$$

$$\frac{p}{\tau_y} = -\frac{2\sigma_{11}+\sigma_{33}}{3\tau_y} = -\frac{\sigma_{33(r)}}{\tau_y} - \frac{2}{3}\sqrt{3}\sqrt{1-\left(\frac{2z}{h}\right)^2}.  \tag{S20}$$

The difference with Prandtl's solution is in multiplier $\sqrt{3}$ instead of 2 in Eq. (S19) for $\sigma_{11}$. The reason is that we use the von Mises condition and $\sigma_{11} = \sigma_{22}$, which results in Eq. (S15), while in Prandtl's solution, the Tresca condition along with plane strain assumption leads to the yield condition $(\sigma_{11} - \sigma_{33})^2 + 4\tau_{31}^2 = \sigma_y^2 = 4\tau_y^2$.

Eq. (S19) and Eq. (S20) lead to the relationship:

$$\frac{\sigma_{11}}{\tau_y} = -\frac{p}{\tau_y} + \frac{\sqrt{3}}{3}\sqrt{1-\left(\frac{2z}{h}\right)^2}.  \tag{S21}$$

Stress $\bar{\sigma}_{11}$ and pressure $\bar{p}$, averaged over the sample thickness are

$$\frac{\bar{\sigma}_{11}}{\tau_y(\bar{p})} = \frac{1}{h}\int_0^h \frac{\sigma_{11}}{\tau_y} dz = \frac{\sigma_{33}(0)}{\tau_y(\bar{p})} + \frac{2r}{h} + \frac{\sqrt{3}\pi}{4} = \frac{\sigma_{33}}{\tau_y(\bar{p})} + \frac{\sqrt{3}\pi}{4};  \tag{S22}$$

$$\frac{\bar{p}}{\tau_y(\bar{p})} = -\frac{\sigma_{33}}{\tau_y(\bar{p})} - \frac{\sqrt{3}\pi}{6}.  \tag{S23}$$

We assumed that $\tau_y$ is constant during averaging and then substituted in the result $\tau_y(\bar{p})$. It is possible to avoid this assumption, but the final equations are getting too bulky and not usable analytically for our purposes. Note that the averaged value of $\bar{\sigma}_{11}$ is much closer to the value of $\sigma_{11}(2z/h)$ at the symmetry plane $\sigma_{11}(0)$ than at the contact surface $\sigma_{11}(1)$. For example, $(\sigma_{11}(0) - \sigma_{33})/(\sqrt{3}\tau_y) = 1$, $\sigma_{11}(1) - \sigma_{33} = 0$, and $(\bar{\sigma}_{11} - \sigma_{33})/(\sqrt{3}\tau_y) = 0.79$. Similar, $(p(0) + \sigma_{33})/(2\tau_y/\sqrt{3}) = -1$, $p(1) - \sigma_{33} = 0$, and $(\bar{\sigma}_{11} - \sigma_{33})/(2\tau_y/\sqrt{3}) = -0.79$.

Eq. (S22) and Eq. (S23) lead to the relationship:

$$\frac{\bar{\sigma}_{11}}{\tau_y(\bar{p})} = -\frac{\bar{p}}{\tau_y(\bar{p})} + \frac{\sqrt{3}\pi}{12}.  \tag{S24}$$

We aim to find the relationship between $\bar{\sigma}_{11}$, $\sigma_{11}(0)$, and $\sigma_{11}(1)$. We will use the following identity:

$$\bar{\sigma}_{11} = \sigma_{11}(1)w + \sigma_{11}(0)(1-w); \qquad w := \frac{\bar{\sigma}_{11}-\sigma_{11}(0)}{\sigma_{11}(1)-\sigma_{11}(0)}.  \tag{S25}$$

Where $w$ is treated as the weight factor. Utilizing Eq. (S19) and Eq. (S22), we obtain:

$$w = 1 - \frac{\pi}{4}\frac{\sigma_y(\bar{p})}{\sigma_y(p(0))} = 1 - \frac{\pi}{4}\frac{\sigma_y^0+b\bar{p}}{\sigma_y^0+bp(0)}.  \tag{S26}$$

Similar,

$$\bar{p} = p(1)w + p(0)(1-w); \qquad w = \frac{\bar{p}-p(0)}{p(1)-p(0)}.  \tag{S27}$$

Here we used the same symbol $w$ because from Eq. (S20) and Eq. (S23), it has the same expression (Eq. (S26)) as for $\sigma_{11}$. Also, we obtain from Eq. (S19) and Eq. (S21):

$$\sigma_{11}(1) = -p(1) = \sigma_{33}; \quad \sigma_{11}(0) = -p(0) + \frac{\sqrt{3}}{3}\tau_y(p(0)) = -p(0) + \frac{1}{3}\sigma_y(p(0)); \quad (S28)$$

from Eq. (S20):

$$p(0) = -\sigma_{33} - 1.155\tau_y(p(0)) = -\sigma_{33} - 0.667\sigma_y(p(0)) = p(1) - 0.667\sigma_y(p(0)); \quad (S29)$$

from Eq. (S24):

$$\bar{\sigma}_{11} = -\bar{p} + 0.453\tau_y(\bar{p}) = -\bar{p} + 0.262\sigma_y(\bar{p}) = 0.262\sigma_y^0 + \bar{p}(0.262b - 1). \quad (S30)$$

Elaborating Eq. (S29) with allowing for Eq. (S9):

$$p(0) = p(1) - 0.667\sigma_y(p(0)) = p(1) - 0.667[\sigma_y^0 + bp(0)] \rightarrow p(0) = \frac{p(1) - 0.667\sigma_y^0}{1 + 0.667b}. \quad (S31)$$

Substitution of Eq. (S31) in Eq. (S26) and Eq. (S27) results in:

$$\bar{p} = p(1)w + \frac{p(1) - 0.667\sigma_y^0}{1 + 0.667b}(1 - w); \quad w = 1 - (0.785 + 0.524b)\frac{\sigma_y^0 + b\bar{p}}{\sigma_y^0 + bp(1)}. \quad (S32)$$

Resolving linear equations Eq. (S32) for $w$ and $p(1)$, we obtain:

$$w = \frac{0.411}{1.910 + b}; \quad (S33)$$

$$p(1) = 0.524\sigma_y^0 + (1 + 0.524b)\bar{p}. \quad (S34)$$

Substituting in Eq. (S9) for $\sigma_y(1)$ in Eq. (S34), we obtain:

$$\sigma_y(1) = \sigma_y^0 + bp(1) = (\sigma_y^0 + b\bar{p})(1 + 0.524b) \quad (S35)$$

Substituting Eq. (S30) and Eq. (S35) in Eq. (S11) results in the final equilibrium equation for parameters $\sigma_y^0$ and $b$ from the best fit to experiments:

$$\frac{d\bar{p}}{dr} = -\frac{2}{\sqrt{3}}\frac{1 + 0.524b}{1 - 0.262b}\frac{\sigma_y^0 + b\bar{p}}{h}. \quad (S36)$$

Eq. (S36) is the final mechanical equilibrium equation expressed in terms of measured pressure $\bar{p}$ averaged of the sample thickness, which is used as Eq. (1) in the main text to determine the pressure dependence of the yield strength. It transforms to the known equation (*10-12*) for $b = 0$ only. We want to use data from all four compression stages as a single data set. To do this, we must justify a way to combine all data in a single plot. Eq. (S36) and its solution in Eq. (1) in the main text have the following properties:

(a) Pressure distribution depends on the dimensionless geometric parameter $r/h$ rather than on $r$ and $h$ separately.

(b) Pressure distribution curves for different applied forces and compression can be overlapped by shifting curves along the $r$ axis without changing $\sigma_y(p)$, since change $r \rightarrow r + C$ does not violate Eq. (S36). Indeed, one can choose the same $p_0$ for all curves and choose constant $C$ for each curve such that $\frac{r+C}{h} = const$ is the same for all curves.

These properties are used in Fig. 1C in the main text. Practically, one can choose a fixed $(p_f, r_f)$ point in the $p - r/h$ plane for all curves to pass through. Then the curve that originally passes through the point $(p_f, r_i)$, should be shifted in the positive direction by the distance $(r_f - r_i)/h$, so that the new curve passes through $(p_f, r_f)$. Then we used all the points in the shifted curve in Fig. 1C to find the best fit for Eq. (S36) (or Eq. (1) in the main text).

**Rationales for the evolution of the crystallite size and dislocation density in ω-Zr during the phase transformation**

Small crystallite size in ω-Zr at the beginning of PT is caused by small transformed regions. The growth of the crystallite size in ω-Zr is related to the growth of these regions in the course of PT. Also, as it follows from (*7*) and the current paper, the reduction in the crystallite size of α-Zr reduces the minimum pressure for initiation of the strain-induced PT $p_\varepsilon^d$ and promotes the PT. That is why the smallest crystallites of α-Zr transform first to ω-Zr, then larger grains transform, so the crystallite size in ω-Zr grows during PT. Since ω-Zr is approximately two times stronger than α-Zr, plastic strain is mostly localized in the α-Zr. That is why plastic strain and strain path do not affect the crystallite size and dislocation density in ω-Zr. Reduction in the dislocation density in ω-Zr is caused by the inverse proportion between the dislocation density and the crystallite size following from Eq. (S1) and Eq. (S2).

**Explanation of the existence of outliers in the evolution of the crystallite size and dislocation density in α-Zr during the phase transformation**

As it follows from Fig. 2D and Fig. 3D, the crystallite size of and dislocation density in ω-Zr during the phase transformation are unique functions of the volume fraction of ω-Zr independent of pressure, plastic strain tensor, and its path. Similar dependence is found for α-Zr in Fig. 2C and Fig. 3C, but there are outliers for $0.38<c<0.52$ obtained at the 2 GPa step. Indeed, at the 2 GPa step and in the two-phase region, the crystallite size of α-Zr remains constant while its volume fraction is larger than 0.6 (Fig. 2A and Fig. 2C), same as the steady value before PT. When the volume fraction of α-Zr gradually decreases to 0.48 towards the culet center, the average crystallite size of α-Zr slightly increases to ~60 nm. This is caused by the statistical effect. As it follows from (*7*) and the current paper, the reduction in the crystallite size of α-Zr reduces the minimum pressure for initiation of the strain-induced PT $p_\varepsilon^d$ and promotes the PT. That is why the smallest crystallites of α-Zr transform first to ω-Zr, increasing the average size of the remaining α-Zr crystallites. It is almost non-detectable at large volume fractions of α-Zr but essential at small volume fractions. Also, constant crystallite size is observed for $r>60$ µm, where, due to friction, plastic deformation is much larger than at the central part. This large plastic strain restores the same steady averaged crystallite size by refining large crystallites. At the center, plastic strain is much smaller and insufficient to restore the steady size. At the 3 GPa step, with further reduction in the volume fractions of α-Zr and an increase in plastic strain, these outliers disappear, and all points belong to the single red curve in Fig. 2C versus volume fractions of α-Zr. Reduction in crystallite size is related to dividing α-Zr crystallite into two or more parts due to PT inside of grains.

A similar statistical effect can explain outliers in the dislocation density in α-Zr for $0.38<c<0.52$ obtained at 2 GPa step. Formally, it is caused by the inverse proportion between the dislocation density and the crystallite size that follows from Eq. (S1) and Eq. (S2). Physically, PT starts and occurs first in the grains with the largest dislocation density, where the probability of strong stress concentrators is higher. Transformation of these grains of α-Zr to ω-Zr decreases the averaged dislocation density in the remaining α-Zr crystallites. It is almost non-detectable at large volume fractions of α-Zr but essential at decreasing volume fractions. Also, constant dislocation density is observed for $r>60$ µm, where plastic deformation is much larger than at the central part. This large plastic strain restores the same steady averaged dislocation density in the large grains. At the center, plastic strain is much smaller and insufficient for restoring the steady dislocation density. At the 3 GPa step, with further reduction in the volume fraction of α-Zr and an increase in plastic strain, these outliers disappear, and all points belong to the single red curve in Fig. 3C versus the volume fraction of ω-Zr. An increase in averaged dislocation density is probably caused by increased dislocation density near new α- ω interfaces to accommodate local transformation strain and decrease crystallite size. Large scatter in both crystallite size and dislocation density in α-Zr near completion of PT is caused by increasing measurement error for a tiny amount of α-Zr.

**Scatter in crystallite size and dislocation density in ω-Zr after completing phase transformation**

While the crystallite size and the dislocation density in ω-Zr after completing the phase transformation are independent of the radius (Fig. 2B and Fig. 3B), there are some scatters around the average along the radius. Also, the dislocation densities are slightly varying between 6, 10, and 14 GPa steps. These scatters cannot be attributed to the dependence of the crystallite size and dislocation density on pressure, plastic strain, and strain path. Indeed, pressure strongly and monotonously reduces, plastic strain strongly and monotonously increases along the radius, and the plastic strain path also changes monotonically. However, there are no clear radial dependence of the crystallite size and the dislocation density. Because of the large fluctuation, the slight difference in the average dislocation density between 6, 10, and 14 GPa steps also cannot be solely attributed to the growing pressure and plastic strain. A possibility is that the observed fluctuations in the crystallite size and the dislocation density after PT completed are due to evolving texture (i.e., dynamically changing distribution of crystallographic orientations and uncharacterized preferred orientations) during the plastic deformation with increasing pressure and errors in post-processing of XRD patterns as described at the end of the section "Dislocation density estimation."

**New findings relative to the previous works**

The effects of severe plastic deformations under high pressure on phase transformations and microstructure evolution are mostly studied with high-pressure torsion (HPT) with metallic or ceramic anvils, see reviews (*8,33-35*). Stationary states after severe plastic straining in terms of torque, hardness, and grain size are well-known in literature, particularly after HPT, along with many cases where they were not observed. However, all these results were not observed in situ but obtained postmortem after pressure release and further treatment during sample preparation for mechanical and structural studies. The direct effect of pressure and the combined effect of pressure and plastic straining on the yield strength, crystallite size, and dislocation density were not determined in the literature. This is very important because, e.g., the yield strength of the ω-Zr doubles at ~13 GPa. During unloading after compression or HPT, additional plastic deformation may occur, which may also cause direct or reverse PT (*36, 37*). Also, several PTs may occur during the loading and others during unloading, e.g., Si-I→Si-II→ Si-XI→ Si-V during loading and Si-V→Si XII & III during unloading (*38, 39*), and the final product does not characterize any PT and processes during the loading. Since after severe plastic deformation a material becomes brittle and internal tensile stresses are present in some regions, damage may also occur.

Moreover, during machining, polishing, and electropolishing of the recovered sample, with or without acids, direct or reverse PT may occur as well, in particular, for Zr (*40*). It was obtained in (*40*) that the grain/crystallite size of ω-Zr is smaller than those of α-Zr, while our in-situ experiments show the opposite. Some samples were characterized six months later than HPT was performed (*41*), and some heterogeneities in hardness distribution along the radius were found. β-Zr was found in (*18*) after compression of optimally oriented highly textured Zr at 1 GPa and after five anvil rotations at 0.5 GPa in (*42*) from the same Zr sample we are using here. However, in situ, we did not find any traces of β-Zr even at 13 GPa. Our results are consistent with the first-principles simulations in (*43*), in which β-Zr exhibits imaginary phonon frequencies and is dynamically unstable at a pressure lower than 25 GPa. Grain size and dislocation density may also change through recovery and recrystallization processes. Thus, in comparison with our in-situ examination, various inaccuracies are introduced in postmortem studies. In addition, pressure during compression and HPT with metallic/ceramic anvils is determined as a total force over total area, which may underestimate the maximum pressure in a sample by a factor of 3 or more (*44, 45*). In particular, the above numbers for PT pressure and corresponding numbers in (*18, 40-42*) for α-ω PT should be multiplied by these correcting factors.

Because of the above problems, the time-dependence of plastic strain-induced PT kinetics was not reported previously and could be reliably determined only in in-situ experiments. The same is true for the minimum pressure for the direct strain-induced PT and, consequently, for the findings that it is independent of the preliminary plastic straining (above some critical magnitude), pressure-accumulated plastic strain, and entire plastic strain path. The existence of the unique curves for the dislocation density and crystallite size for both α-Zr and ω-Zr during α-ω PT, independent of pressure, $\boldsymbol{\varepsilon}_p$ and $\boldsymbol{\varepsilon}_p^{path}$ has not been reported in the previous studies.

Our results about the existence of multiple steady states are consistent with known results that different ways to produce severe plastic deformation (e.g., HPT, equal channel extrusion, ball milling, etc.) lead to different steady grain sizes (*8, 33, 34*). However, our results also find that the different steady states in terms of the crystallite/grain size, dislocation density, and the minimum pressure for the strain-induced PT can be produced in the same device by the same method just by increasing the height of asperities and, consequently, the contact friction. Being different from the previous studies, the existence of the multiple steady states is proved in situ under high pressure in our study.

Our result in Fig. 1A on the existence of the fixed isotropic pressure-dependent surface of perfect plasticity independent of $\boldsymbol{\varepsilon}_p$ and $\boldsymbol{\varepsilon}_p^{path}$ is far beyond the existence of the steady hardness, the same for different processing techniques and initial states. An important point is how to relate this surface with the traditional evolving yield surface, which is anisotropic and depends on $\boldsymbol{\varepsilon}_p$ and $\boldsymbol{\varepsilon}_p^{path}$. In addition, our finding is formulated in the language of plasticity theory (plastic strain and strain path tensors, yield surface, etc.) instead of technological language, which allows one to use the obtained knowledge to significantly enrich fundamental plasticity in the formulation and application of plastic models and computer simulations of various processes. Note that the isotropy of the surface of perfect plasticity $\varphi(\boldsymbol{s}) = \sigma_y(p)$ follows not only from experiments but from the theory. Indeed, since initially polycrystalline material with stochastic grain orientation without texture is isotropic, its anisotropy during deformation can come from $\boldsymbol{\varepsilon}_p$ and $\boldsymbol{\varepsilon}_p^{path}$ only, i.e., it is strain-induced. Since $\varphi(\boldsymbol{s}) = \sigma_y(p)$ is independent of $\boldsymbol{\varepsilon}_p$ and $\boldsymbol{\varepsilon}_p^{path}$, the only source for anisotropy disappears.

Similarly, the existence of (a) the steady crystallite/grain size and dislocation density determined in situ under high pressure and independent of pressure, $\boldsymbol{\varepsilon}_p$, and $\boldsymbol{\varepsilon}_p^{path}$, and (b) its connection to the surface of perfect plasticity and the minimum pressure for the direct strain-induced PT, both independent of $\boldsymbol{\varepsilon}_p$ and $\boldsymbol{\varepsilon}_p^{path}$, is well beyond of the known postmortem finding of the steady grain size and (in few cases) the dislocation density, the same for different processing techniques and initial states. Note that the steady state in the yield strength does not correspond to the steady state in torque in high-pressure torsion (*45*), mostly due to the complexity of the friction condition.

**Rationales for the reduction in the minimum pressure for the strain-induced PT with decreasing crystallite size and increasing dislocation density**

As suggested in our analytical model (*5*) and phase field models (*46, 47*), plastic strain-induced PT occurs by nucleation at the tip of a dislocation pileup as the strongest possible stress concentrator. All components of stress tensor $\boldsymbol{\sigma}$ at the tip of dislocation pileup, modeled as a superdislocation, are:
$$\boldsymbol{\sigma} \sim \tau l \sim N \qquad (S37)$$
where $\tau$ is the applied shear stress limited by the yield strength in shear $\tau_y$, $l$ is the length of the dislocation pileup, and $N$ is the number of dislocations in a pileup. The higher the dislocation density, the higher the probability of the appearance of dislocation pileups with a larger number of dislocations. This trivially explains reducing the minimum pressure for the strain-induced PT with increasing dislocation density. However, since $l$ is limited by the fraction of the grain size (e.g., half of the grain size), the main conclusion in (*5*) was that the greater grain size is the stronger reduction in the PT pressure, i.e., opposite to what we found in experiments. Our later phase field (*46,47*), molecular dynamics (*48*), and concurrent atomistic-continuum simulations (*49*) allow us to resolve the problem, at least qualitatively. In contrast to the analytical solution utilized in (*5*), $l$ is not related to the grain size since most dislocations are localized at the grain boundary producing a step (superdislocation, Fig. S5) with effective length $l=Nb<<d$, where $b$ is the magnitude of the Burgers vector. At the same time, $\tau = \tau_y$ increases with the decrease in $d$ according to the Hall-Petch relationship $\tau_y = \tau_0 + kd^{-0.5}$, where $\tau_0$ and $k$ are material parameters. That is why the minimum pressure for the strain-induced PT decreases with decreasing crystallite size.

**On the possible source of the time dependence of the kinetics of strain-induced PTs**

It was generally accepted that during shear under high pressure, PT stops when shear stops (*5-8, 50*). That means that time is not a governing parameter and plastic strain plays a role of a time-like parameter. A nanoscale rationale in (*5*) explaining this statement was that barrierless nucleation at the tip of the dislocation pileup occurs extremely fast and, since stress decreases like *1/r* with distance from the tip *r*, grows is very limited and is arrested when phase interface is equilibrated. Since this process occurs in a much shorter time than the measurement time, a time-dependent component is not detectable, and plastic strain is the only governing parameter. This was implemented in (*5*) in the strain-controlled kinetic equation, see Eq. (2) in the main text. This equation was confirmed by experiments in (*7*), but the time-dependent component of the kinetics at fixed load/torque was not checked because it was not expected. After we found here the time dependence of the kinetics of strain-induced PT experimentally, we can revisit the results of the phase-field simulations to rationalize it. Fig. S6 shows the time evolution of the phase and dislocation structures at the fixed applied normal stress and shear strain after phase nucleation in the right grain at the dislocation pileup in the left grain. Applied normal stress is 10 times lower than the PT pressure under hydrostatic conditions. One can see that after nucleation, the high-pressure phase significantly grows and reaches the opposite grain boundary, the number of dislocations in the dislocation pileup in the left grain increases (especially within step at the grain boundary), dislocations nucleate and evolve in the right grain, the second nucleus appears at the dislocation pileup that develops within the right grain, then nuclei coalesce, and the stationary phase and dislocation configurations is achieved. The time scale for phase and dislocation evolution is determined by two kinetic coefficients, which are different for different materials. If the first measurement at the material point in a sample in DAC completes before a stationary state is reached, this evolution is undetectable, the entire process looks instantaneous, and the kinetics of the PT is fully plastic strain controlled. In the opposite case, phase evolution at the fixed strain will be observed and time-dependent component of the kinetics should be characterized and formalized.

**Supplementary figures**

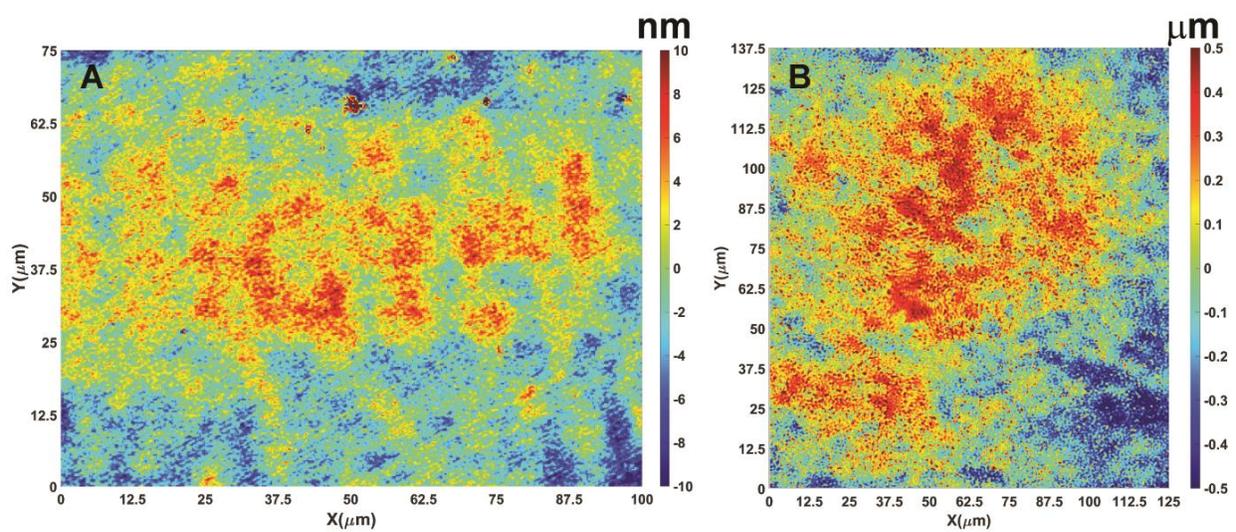

**Fig. S1. Surface asperity profile of a smooth anvil and a rough-DA.** (a) a traditional smooth diamond anvil with a range [-10 nm; 10 nm] and (b) a rough-polished diamond anvil (rough-DA) with a range [-500 nm; 500 nm].

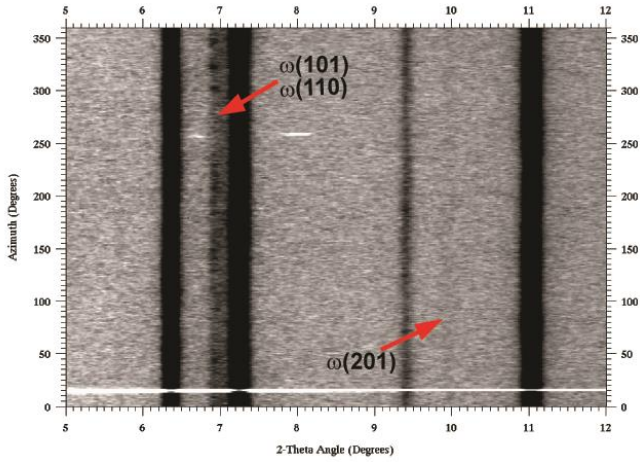

**Fig. S2. Unrolled diffraction image of Zr when ω-Zr first emerged at 0.67 GPa at culet center.**

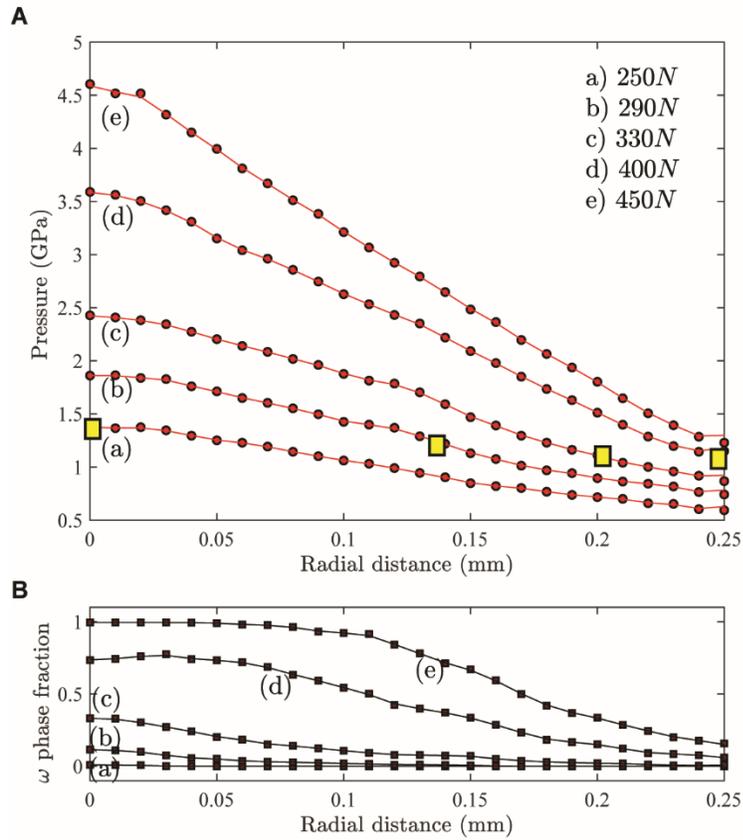

**Fig. S3. Radial distribution of (A) α-Zr pressure and (B) ω-Zr volume fraction in a sample deformed with smooth anvils.** Different applied forces represent different compression stages. Yellow squares show the minimum PT pressure $p_\varepsilon^d$ =1.36 GPa at different compression stages and at different radii where ω-Zr was first observed. Since plastic strain, plastic strain path, and pressure-strain path are very different at different locations and compression stages and $p_\varepsilon^d$ is independent of the locations, then $p_\varepsilon^d$ is independent of plastic strain, plastic strain path, and pressure-strain path.

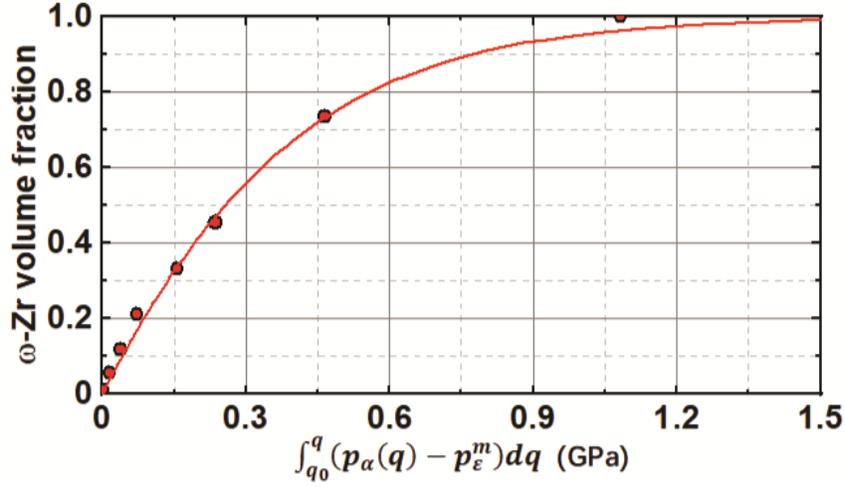

**Fig. S4. Kinetics of α-ω phase transformation in Zr with smooth diamond anvil.** Compared to rough-DA experiment, kinetics shows different nonlinear features corresponding to the first-order reaction with parameters $a=1$, $k=11.65$, and $B=1.35$ in Eq. (2) $\frac{dc}{dq} = k \frac{B(1-c)^a}{B(1-c)+c} \left( \frac{p_\alpha(q) - p_\varepsilon^d}{p_h^d - p_\varepsilon^d} \right)$, instead of $a=l=0$, $B=1$ for the experiment with rough-DA.

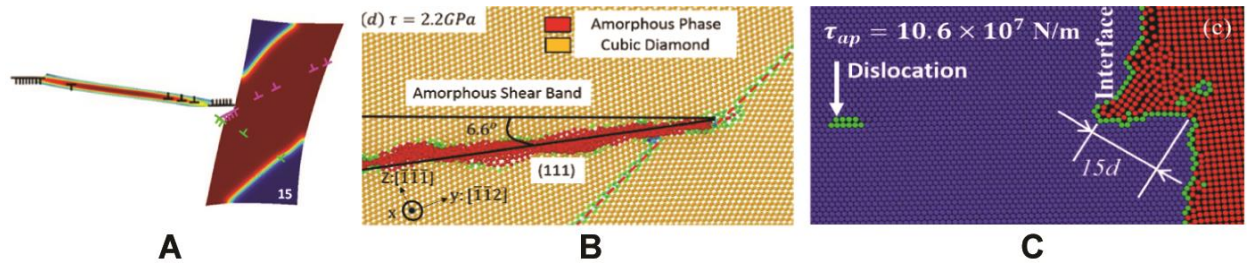

**Fig. S5. Dislocation pileups producing a step at the grain boundary or phase interface that causes a phase transformation.** (**A**) Dislocation pileup in the left grain produces step at the grain boundary and cubic to tetragonal PT and dislocation slip in the right grain. Phase-field approach results from (*47*). (**B**) Dislocation pileup in the right grain produces a step at the grain boundary in Si I and amorphization in the left grain. Molecular dynamics results from (*48*). (**C**) Step at the phase interface boundary consisting of 15 dislocations and causing cubic to hexagonal PT. The atomistic portion of the concurrent continuum-atomistic approach from (*49*). Adopted with changes from (*47-49*) with permissions.

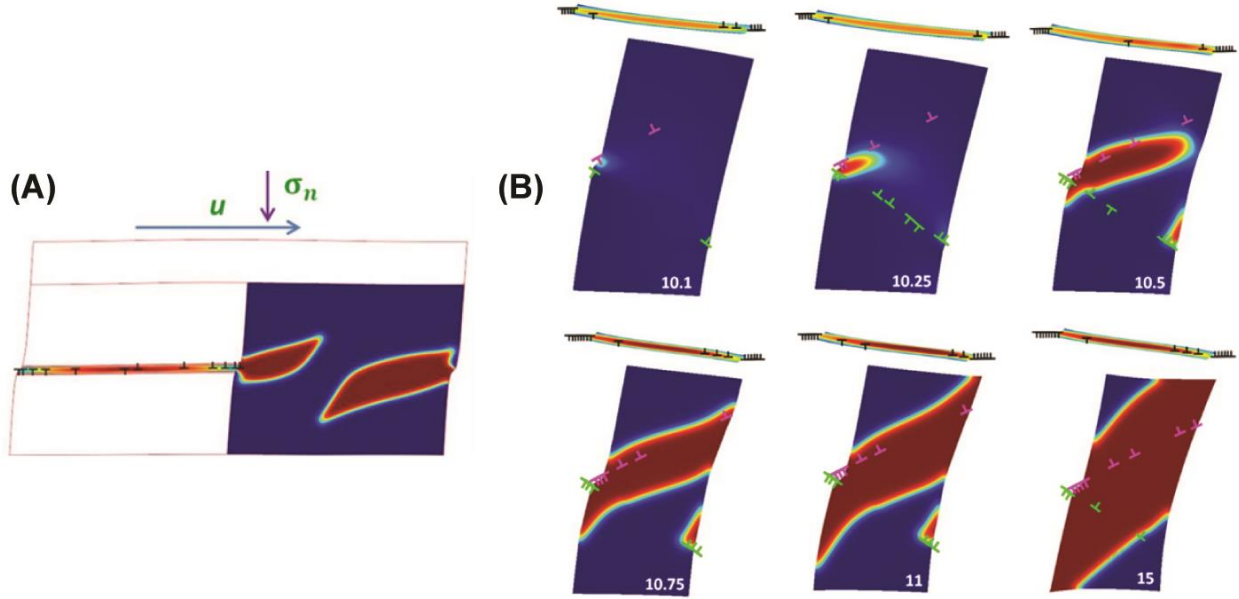

**Fig. S6. Time evolution of the phase and dislocation structures at the fixed applied normal stress and shear strain after phase nucleation in the right grain at the tip of dislocation pileup in the left grain.** (**A**) Schematics of grains with an initial solution for dislocation pileup and nucleated high-pressure phase (red) (*46*). (**B**) Nucleation and growth of the high-pressure phase (red) in the right grain caused by an evolving dislocation pileup in the left grain, which is shown at the top of each right grain (*47*). Results are obtained with the phase-field approach. Adopted with changes from (*46, 47*) with permission.

**Supplementary tables**

**Table S1.** Parameters of 3$^{rd}$ Birch-Murnaghan equation of state of Zr used in this study.

| Zr phase | $V_0$ (per formula unit) | $K_0$ | $K_0'$ |
|---|---|---|---|
| α-Zr | 23.272 Å$^3$ | 92.2 GPa | 3.43 |
| ω-Zr | 22.870 Å$^3$ | 102.4 GPa | 2.93 |

**Table S2.** The thickness of Zr sample with rough-DA in this study at corresponding compression step. 0.67 GPa corresponds to the step when ω-Zr first emerge.

| Compression step | initial | 0.67 GPa | 2 GPa | 3 GPa | 6 GPa | 10 GPa | 14 GPa |
|---|---|---|---|---|---|---|---|
| Thickness (μm) | 163 | 101 | 56 | 48 | 40 | 32 | 26 |